\documentclass{aa}
  \usepackage[T1]{fontenc}
  \usepackage[varg]{txfonts}
  \usepackage{graphicx}
  \usepackage{natbib}
  \usepackage{array}
  \usepackage{xspace}

\usepackage{subfigure} 

\def\BE{\begin{equation}}
\def\EE{\end{equation}}
\def\BA{\begin{array}}
\def\EA{\end{array}}
\def\BAN{\begin{eqnarray}}
\def\EAN{\end{eqnarray}}

\def\FIG #1 #2 [#3] #4\par{%
 \begin{figure}
   \resizebox{\hsize}{!}{\includegraphics[#3]{#2}}
   \caption{#4}
    \label{#1}
 \end{figure}
}

\def\FIGG #1 #2 #3 [#4] #5\par{%
 \begin{figure}[!h]
   \begin{center}
   \includegraphics*[#4]{#2}
   \includegraphics*[#4]{#3}
   \caption{\label{#1}#5}
   \end{center}
 \end{figure}
}

\def\FIGs #1 #2 #3 #4 #5 [#6] #7\par{%
 \begin{figure}[!h]
   \begin{center}
       \includegraphics*[#6]{#2}
       \includegraphics*[#6]{#3}
       \includegraphics*[#6]{#4}
       \includegraphics*[#6]{#5}
       \caption{\label{#1}#7}
   \end{center}
 \end{figure}
}

\def\FIGss #1 #2 #3 #4 #5 #6 #7 [#8] #9\par{%
 \begin{figure}[!h]
   \begin{center}
       \includegraphics*[#8]{#2}
       \includegraphics*[#8]{#3}
       \includegraphics*[#8]{#4}
       \includegraphics*[#8]{#5}
       \includegraphics*[#8]{#6}
       \includegraphics*[#8]{#7}
       \caption{\label{#1}#9}
   \end{center}
 \end{figure}
}

\begin{document}

\title{Thermonuclear explosions of rapidly rotating white dwarfs -- 
	II. Detonations} 
\titlerunning{Prompt detonations of rapidly rotating white dwarfs}

\author{J.M.M.~Pfannes\inst{1}, J.C.~Niemeyer\inst{1,2} and W.~Schmidt\inst{1,2}}
 \authorrunning{J.M.M.~Pfannes, J.C.~Niemeyer and W.~Schmidt}         
   \offprints{J.C.~Niemeyer}

\titlerunning{Detonations of rapidly rotating white dwarfs}
\authorrunning{J.M.M.~Pfannes et al.}

\institute{Lehrstuhl f\"ur Astronomie,
	Universit\"at W\"urzburg, Am Hubland, D-97074 W\"urzburg, Germany \\
	\email{pfannes@astro.physik.uni-wuerzburg.de}
	\and
	Institut f\"ur Astrophysik, Universit\"at G\"ottingen, Friedrich-Hund-Platz 1,
D-37077 G\"ottingen, Germany \\
	\email{[niemeyer;schmidt]@astro.physik.uni-goettingen.de}
}

   \date{the date of receipt and acceptance should be inserted later}

   \abstract
       % context
       {Superluminous type Ia supernovae (SNe Ia) may be explained
       by super-Chandrasekhar-mass explosions of rapidly rotating white dwarfs
       (WDs). In a preceding paper, we showed that the deflagration scenario 
       applied to rapidly rotating WDs generates explosions that
       cannot explain the majority of SNe Ia.}   
       %
       % aims
       {Rotation of the progenitor star allows super-Chandrasekhar mass
       WDs to form that have a shallower density stratification. We use simple
       estimates of the production of intermediate and iron group
       elements in pure detonations of rapidly rotating WDs
       to assess their viability in explaining rare SNe Ia.}
       %
       % methods
       {We numerically construct WDs in
       hydrostatic equilibrium that rotate according to a variety of rotation
       laws. The explosion products are estimated by considering the density
       stratification and by evaluating the result of hydrodynamics
       simulations.}
       %
       % results
       {We show that a significant amount of intermediate mass
       elements is produced for
       theoretically motivated rotation laws,
       even for prompt
       detonations of WDs.}
       %
       % conclusions
       {Rapidly rotating WDs that detonate may provide an explanation of
         rare superluminous SNe Ia 
       in terms of both burning species and explosion kinematics.}

 
\maketitle

%%%%%%%%%%%%%%%%%%%%%%%%%%%%%%%%%%%%%%%%%%%%%%%%%%%%%%%%%%%%%%%%%%%%%%%%%
%%%%%%%%%%%%%%%%%%%%%%%%%%%%%%%%%%%%%%%%%%%%%%%%%%%%%%%%%%%%%%%%%%%%%%%%%
%%%%%%%%%%%%%%%%%%%%%%%%%%%%%%%%%%%%%%%%%%%%%%%%%%%%%%%%%%%%%%%%%%%%%%%%%

\section{Introduction}
\label{intro}

Type Ia supernovae (SNe Ia) were long believed to form a relatively
homogeneous class of events in terms of their spectra, peak
luminosities, and light curves. Surveys, however, have shown
that the distribution of SNe Ia properties is substantially broader
than previously anticipated, including those of highly peculiar events such as
SN 2007ax \citep{Kasliwal_etal}, SN 2005hk \citep{Stanishev_etal}, and
SN 2003fg \citep{Howell_etal}. The latter, in particular, has been
interpreted in terms of a super-Chandrasekhar-mass explosion of a
rapidly rotating white dwarf (WD) \citep{Jeffery_etal}.

Motivated by the goal of explaining the presumed homogeneity of
SNe Ia within a single explosion scenario, most multi-dimensional hydrodynamical
simulations so far have focused on turbulent deflagrations
\citep{2006A&A...446..627S,snob} or 
delayed detonations \citep{GamKhok05,RoepkeNiemeyer07} of
Chandrasekhar-mass WDs (this class includes
the gravitationally confined detonation model proposed by
\citet{GCDmodel}). The delayed detonation model yields an explosion consistent with
kinematics as well as spectra, light curves, and nucleosynthesis of type Ia supernovae. 
So far, a major disadvantage of this model has been
that the conditions for a deflagration-to-detonation transition (DDT) as
proposed by \citet{1991A&A...245..114K} and \citet{WW94}, are
questionable in the context of WD matter
\citep{1999ApJ...523L..57N}. Although 
\citet{2008ApJ...681..470P} proposed a theoretical
explanation of delayed detonations, the numerical studies by \citet{WoosKer09} and \citet{SchmCir09} nevertheless impose severe constraints on DDTs. 
Prompt detonations, on the other hand, have been 
considered to be infeasible for explaining SNe Ia since the pioneering work of 
\citet{Arn69}, as too little material is burned at sufficiently low
densities to produce intermediate-mass elements (IMEs). 

The opposite problem occurs in the case of rapidly rotating WDs, which
burn as turbulent deflagrations (\citet{PaperI}, hereafter referred to as Paper~I). Here, too much stellar material remains
unburned as a consequence of the lower density at large radii. Combining the fast
propagation of a detonation front with the shallower stratification of
the stellar fuel, 
rapidly rotating initial models have already been proposed as a possible means
of repairing the prompt
detonation mechanism by \citet{1992A&A...254..177S}. However, in their models
the bulk of WD matter was nevertheless burnt to form iron-group elements (IGEs) and, accordingly, even
rapid rotation did not change the situation. Whenever a significant amount of
IMEs was produced by rotation in their study, the amount of IGEs was
simultaneously too high.

In this work, we revisit the scenario of promptly detonating carbon-oxygen (CO)
WDs. In contrast to \citet{1992A&A...254..177S}, we investigate a
variety of rotation laws including differentially rotating models
inspired by the results of \citet{2005A&A...435..967Y}. Some of the models in
our sample significantly exceed the Chandrasekhar mass, which is indeed
suggested by some observations, for instance, \citet{Howell_etal}. 
For this reason, we do not attempt
to interpret our results as those for normal SNe Ia but instead look for
possible connections with observed peculiar supernovae. 

It is important to emphasize that we only attempt to predict the
expected range of masses of the produced IMEs and IGEs, to determine whether these types of explosions might account
for rare SN Ia events. To this end, we first provide two simple estimates based
on the equilibrium stratification of the WD and different choices of
the density threshold for burning into nuclear statistical equilibrium
(NSE). These numbers can be interpreted as plausible upper and lower
bounds on the produced IME masses in these events. This interpretation
is confirmed by the nucleosynthetic yields obtained by post-processing
a full explosion simulation, which lie in between these bounds. Going
beyond these estimates would require a parameter study of rotation
laws and ignition conditions of sufficient resolution to capture the
detonation front. Given the speculative nature of this model, we do
not think that this effort is warranted at present.

Based on these simplifying assumptions, we find that 
a significant amount of IMEs 
(0.1 to 0.4~$M_{\sun}$) is being produced  
accompanied by a Super-Chandrasekhar mass amount of IGEs 
(1.5 to 1.8~$M_{\sun}$)
and a marginal amount of unburnt stellar material.
The ejecta expand at higher radial velocities because of the greater amount
of nuclear energy released as a consequence of the detonation. 
Unburnt stellar material is leftover only in stellar regions
close to the stellar surface at the equatorial plane, whereas iron group
elements are 
predominant within the star and at the stellar poles as a consequence of
the density stratification that is shaped by rapid rotation. For the same
reason, a bulge of IMEs is generated in the outer regions of the
equatorial plane.

The discussion will proceed as follows. 
In Sect. 2, we introduce our numerical method. 
The procedure that allows us to
estimate the composition of burning species is
discussed in 
Sect. 3. The process of a typical prompt detonation initiated within
a rapidly rotating WD is described in Sect. 4. Section 5 contains an
interpretation with respect to spectral features. The detailed stellar
composition is investigated in Sect. 6. Section 7 concludes the paper.

%%%%%%%%%%%%%%%%%%%%%%%%%%%%%%%%%%%%%%%%%%%%%%%%%%%%%%%%%%%%%%%%%%%%%%%%%
%%%%%%%%%%%%%%%%%%%%%%%%%%%%%%%%%%%%%%%%%%%%%%%%%%%%%%%%%%%%%%%%%%%%%%%%%
%%%%%%%%%%%%%%%%%%%%%%%%%%%%%%%%%%%%%%%%%%%%%%%%%%%%%%%%%%%%%%%%%%%%%%%%%

\section{Method}

To study the outcome of detonations of rapidly rotating WDs, we
constructed rotators with different rotation laws that 
can potentially give rise to SN~Ia progenitors.

The WDs were assumed to consist of carbon and oxygen in equal proportions 
with central densities of~$2.0\times10^9$~g/cm$^3$. 
Depending on the detailed rotation law, their masses ranged from 1.4 to
2.1~M$_{\sun}$ (cf. Fig.~\ref{fig:22AWD3_densty}). 
The equation of state considered the thermodynamical properties of a
photon, electron, and baryonic gas, as well as
electron-positron-pair-creation. The temperature profile was adjusted to
describe the
evolution of the accreting WD (cf. the study of \citet{2005A&A...435..967Y}). 
Our simulations were carried out in three spatial dimensions. We used a
moving equidistant cartesian grid with $100^3$~cells (initial grid
spacing of $1.0\times10^7$~cm) for the computation of a full star explosion. The
detonation was ignited centrally for most of the calculations.
Gravitational effects were taken into account by using a multipole solver  
(see Paper~I for more details of the rotation laws and the
numerical implementation). 

Since the detonation front incinerates the star before changes
in the stellar structure can occur, an estimate
of the burning products of detonating WDs can be obtained 
by simply looking at the material densities present in the 
hydrostatic case. 

The thermal and chemical structure of {\it planar} thermonuclear
detonations is well understood \citep{ImshennikKhokhlov1984} and
yields a transition density for burning into NSE of 
approximately $10^7$ g/cm$^3$. However, in the models that we consider, the
detonation front is subject to strong shear, which, in combination with
the cellular front instability \citep{Gamezoetal1999,Timmesetal2000} can be expected
to give rise to strong local variations in the temperature and density
profile. Most of these effects have a tendency to lower the
burning temperature at fixed density and thus postpone the burning to
NSE to a higher threshold. A full exploration of these phenomena is beyond 
the scope of our present study. Instead, we argue that the higher transition density of
planar deflagrations (roughly 5 $\times 10^7$ g/cm$^3$) yields a
plausible upper cutoff, the lower cutoff being given by planar detonations. 

Hence, to estimate the range of masses of IMEs produced by the detonation,
we considered two different burning thresholds for NSE that can be considered as
limiting cases. 
As an upper bound to the transition density, we applied the same density thresholds as
commonly used for deflagrations. There,
within regions of high density ($\rho > 5.248 \times
10^{\,7}~\mathrm{g/cm^3}$), IGEs are assumed to be
produced. Medium density regions ($ 5.248 \times 10^{\,7}~\mathrm{g/cm^3} >
\rho > 1.047 \times 10^7~\mathrm{g/cm^3}$) and low
density regions  ($\rho \leq 1.047 \times 10^7~\mathrm{g/cm^3}$)
yield IMEs and unburnt fuel (``C+O''), respectively. We refer to those
{\it high NSE burning density threshold} values hereafter as the HBT limiting case. 

The transition density for planar detonations was chosen to be a lower
bound ({\it low NSE burning density threshold}, in short LBT). 
IGEs are assumed to be produced for $\rho > 10^{\,7}~\mathrm{g/cm^3}$, IMEs
are obtained for lower densities.

Finally, to take into account the explosion dynamics, hydrodynamical simulations are
necessary.  
Thus, in addition to the analysis of the density stratification, we used the 
hydrodynamics code PROMETHEUS (cf. Paper~I) to simulate prompt detonations
of rotating WDs. By employing the method described by
\citet{2004A&A...425.1029T}, the data from
these simulations were post-processed to derive a more accurate
result for the explosive nucleosynthesis. We emphasize, however, that the
resolution and range of explored parameters are insufficient to
provide a robust investigation of the nuclear yields of this
scenario. 

Burning is realized by means of the level set method,
introduced by \citet{OshSet88} and applied to SN~Ia calculations by
\citet{RHNKG99b} for pure deflagrations. \citet{2005A&A...438..611G}
applied this method to supersonic burning fronts in the context of delayed
detonations. 
In both cases, the flame front is treated as a surface that separates fuel from
ashes with a sharp discontinuity. 
For detonations, the speed of the flame depends on the mass density and is
calculated by a linear interpolation of the values in Fig. 2 of \citet{Sha99}.
For the hydrodynamical simulations we used the HBT limiting case, taking into
account that temperature is underestimated once densities decrease below $ 5.248
\times 10^{\,7}~\mathrm{g/cm^3}$. 

\begin{figure}[t]\resizebox{\hsize}{!}{
\includegraphics[clip]{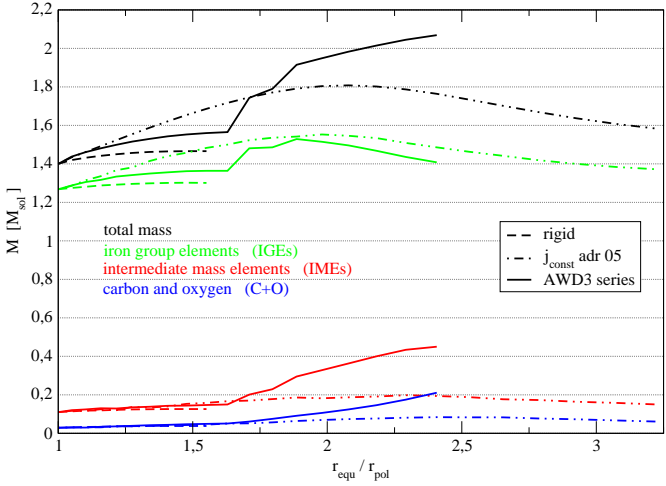}}
\caption{Fractions of the total mass in the high density
  regime ($\rho > 5.248 \times 10^{\,7}~\mathrm{g/cm^3}$, ``IGEs''), the
  medium density regime ($ 5.248 \times 10^{\,7}~\mathrm{g/cm^3} >
  \rho > 1.047 \times 10^7~\mathrm{g/cm^3}$, ``IMEs''), and the low
  density regime ($\rho \leq 1.047 \times 10^7~\mathrm{g/cm^3}$ ,
  ``C+O'') for different rotation laws 
  versus the ratio of equatorial to polar radius (denotation according to the
  HBT limiting case)} 
\label{fig:density_profile_comp}
\end{figure}

\begin{figure*}[th]
  {\includegraphics[width=17cm]{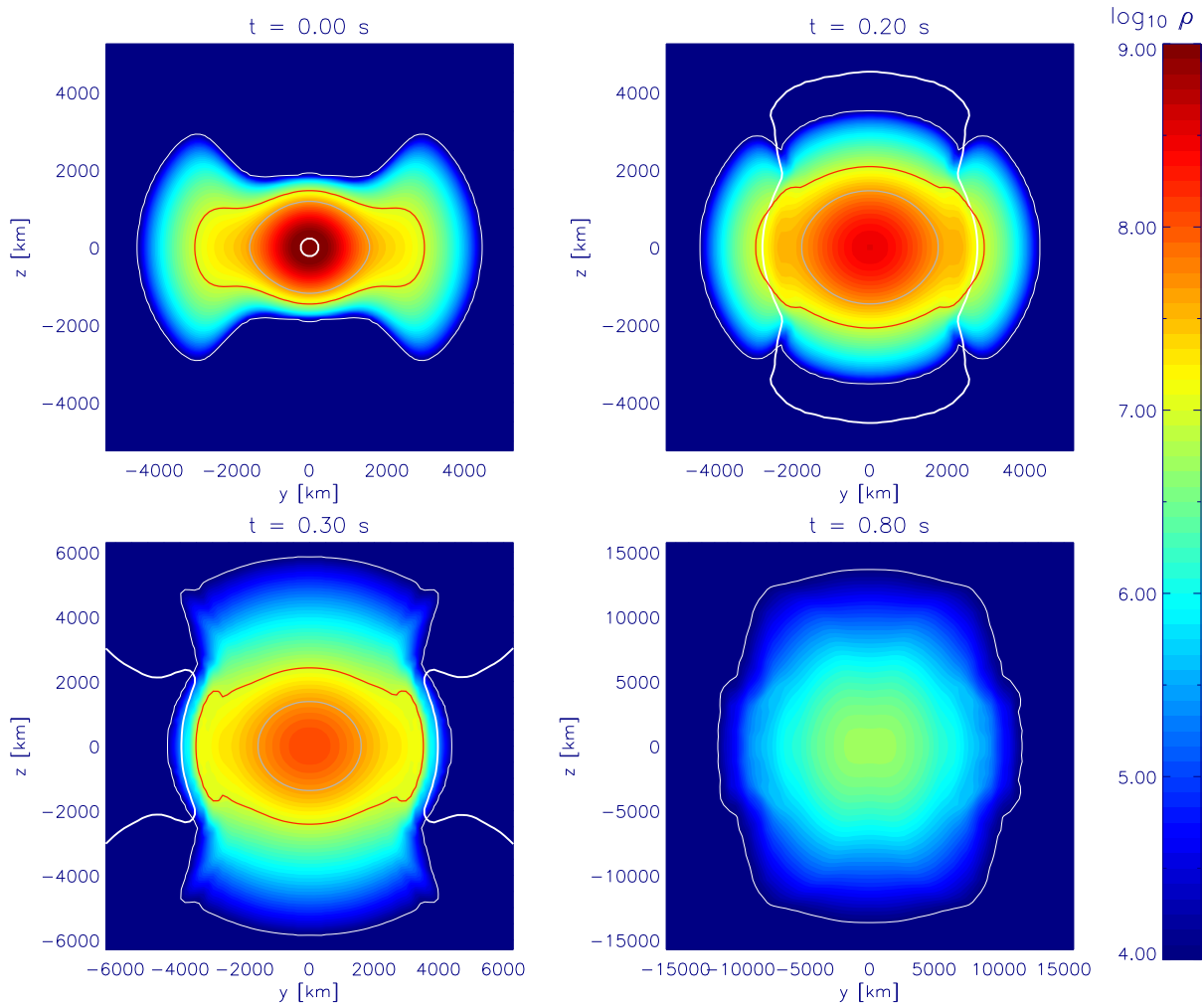}
    }    
  \caption{Density contour lines for the prompt detonation of the
    AWD3 rotator at different instants of time (the white, thick line
    represents the detonation front).
    Cross-sections along the rotation axis of a full star
    simulation are shown.} 
  \label{fig:22AWD3_densty}
\end{figure*}

\section{Estimate of the burning products}

Owing to detonations being able to propagate through an essentially
undisturbed equilibrium configuration, we can
estimate the burning species on the basis of the
hydrostatic density stratification of the progenitor star.
 
While the detonation front crosses dense material, 
$^{56}\mathrm{Ni}$ is mainly produced
(for densities $\rho \gtrsim 3 \times 10^{\,9}~\mathrm{g/cm^3}$, although
adverse electron capturing becomes important). 
A sufficient amount of IMEs can only be obtained if there is
enough material at densities lower than about 
$5 \times 10^7~\mathrm{g/cm^3}$ 
or $1 \times 10^7~\mathrm{g/cm^3}$
(HBT and LBT limiting cases, respectively)
in the progenitor star.
Figure \ref{fig:density_profile_comp} illustrates the mass proportions
in the corresponding density regimes for the rigid rotator, a rotator with
approximately constant angular momentum ($j_{const}$), and a rotator
obeying the rotation law derived for an {\it accreting white dwarf} (AWD;
cf. Paper~I and the study of \citet{2004A&A...419..623Y}). For 
every kind of
rotator, the ratio of the equatorial radius to the polar radius  
steadily increases 
along the x-coordinate
and the WD is deformed into
a doughnut-like shape with increasingly rapid rotation. 

From Fig. \ref{fig:density_profile_comp}, it is clear that for 
the AWD3 series, fewer IGEs and, at the same time, more IMEs are
generated in comparison to the other models. 
The AWD3 rotator with the fraction of the radii $r_{equator} / r_{pol} = 2.4$
contains  
$1.41~M_{\sun}$ of IGEs and $0.45~M_{\sun}$ of IMEs,
or
$1.86~M_{\sun}$ of IGEs and $0.21~M_{\sun}$ of IMEs
(HBT and LBT limiting cases, respectively).
In contrast, the $j_{const}$ sequence used by \citet{1992A&A...254..177S} 
produces more IGEs and fewer IMEs. 

%
%%%%%%%%%%%%%%%%%%%%%%%%%%%%%%%%%%%%%%%%%%%%%%%%%%%%%%%%%%%%%%%%%%%%%%%%%
%%%%%%%%%%%%%%%%%%%%%%%%%%%%%%%%%%%%%%%%%%%%%%%%%%%%%%%%%%%%%%%%%%%%%%%%%
%%%%%%%%%%%%%%%%%%%%%%%%%%%%%%%%%%%%%%%%%%%%%%%%%%%%%%%%%%%%%%%%%%%%%%%%%

\section{Evolution of a prompt detonation}

Figure \ref{fig:22AWD3_densty} shows the density contours of
the centrally ignited ``AWD3 detonation'' explosion model. 
The burning front (white, thick line) moves outwards 
(without stopping after burning has ceased but continues to propagate outside the
star for numerical reasons). The large amount of IMEs resulting from
the prompt detonation of the AWD3 rotator stems from the 
equatorial bulge of low density material that
is typical of accreting WD rotation (the density thresholds for 
the production of IMEs and IGEs, respectively, are indicated by the contour
lines inside the burning front 
for the HBT limiting case).

\begin{table}%[tp]
\begin{center}
\begin{tabular}{|r||*{3}{c|}}
\hline
                            & HBT        & LBT           & postproc \\
\hline\hline            %      PPA        promgrid / tracergrid   postproc
$M_{tot}~[M{\sun}]$         & \multicolumn{2}{|c|} {2.07}    &  2.12    \\

\hline\hline
\multicolumn{4}{|l|} {t~=~0\,s} \\
\hline
$E_{grav}~$[10$^{\,51}$\,erg] &  \multicolumn{2}{|c|} {-4.500}  & \\
\hline
$E_{rot}~$[10$^{\,50}$\,erg]  & \multicolumn{2}{|c|} {4.55} & \\
\hline
$\beta~[\%]$                  & \multicolumn{2}{|c|} {10.11}  & \\
\hline\hline
\multicolumn{4}{|l|} {t~=~10\,s} \\
\hline
$E_{kin}~$[10$^{\,50}$\,erg]  &  15.88 & & \\
\hline
$E_{tot}~$[10$^{\,50}$\,erg]  & 15.30 & & \\
\hline
$E_{nuc}~$[10$^{\,51}$\,erg]  &  2.650 & & \\
\hline
IGEs~$[M{\sun}]$         & 1.41 & 1.86  & 1.75 \\
$[\%\, M_{tot}]$         & 68   & 90 &   83 \\
\hline
IMEs~$[M{\sun}]$         & 0.45 & 0.21 & 0.29 \\
$[\%\, M_{tot}]$         & 22   & 10  &  14  \\
\hline
C+O~$[M{\sun}]$          & 0.21 & 0 & 0.08 \\
$[\%\, M_{tot}]$         & 10   & 0    &  3   \\
\hline
\end{tabular}
\end{center}
\caption{Quantities 
  for the ``AWD3 detonation'' explosion model from
  the initial model in the HBT and LBT limiting cases.}
\label{tab:yields_mod22}
\end{table}

Table \ref{tab:yields_mod22} summarizes the values derived
directly from the rotating initial model for both HBT and LBT limiting cases
(left and middle column respectively). The ratio of 
  rotational to gravitational energy of the progenitor star
is denoted by $\beta$.
The energies are obtained from the hydrodynamical simulation. 
Moreover, the results
from post-processing of the simulation data are listed
(cf.~Sect.~\ref{sc:post}).  
Remarkably, the results of the post-processed simulation are marked on either side
by the limiting case estimates of the equilibrium model,
confirming the consistency of our assumptions. 

The mass of the
tracer particles used in post-processing the simulation data
\citep{Rei01} adds up to $2.12~M_{\sun}$, which exceeds the stellar
mass of the AWD3 rotator by more than $2~\%$. 
This is a consequence of the
small number of tracer particles, namely
$15^{\,3}$, employed by the post-processing, which is 
computationally demanding because of the high detonation temperatures. 

\begin{figure}[th]
  \begin{minipage}{0.5\textwidth}
    \centering
    \subfigure[``AWD3 deflagration $C_e = 5 \times 10^{\,3}\,$''] 
    {\includegraphics[width=\linewidth]{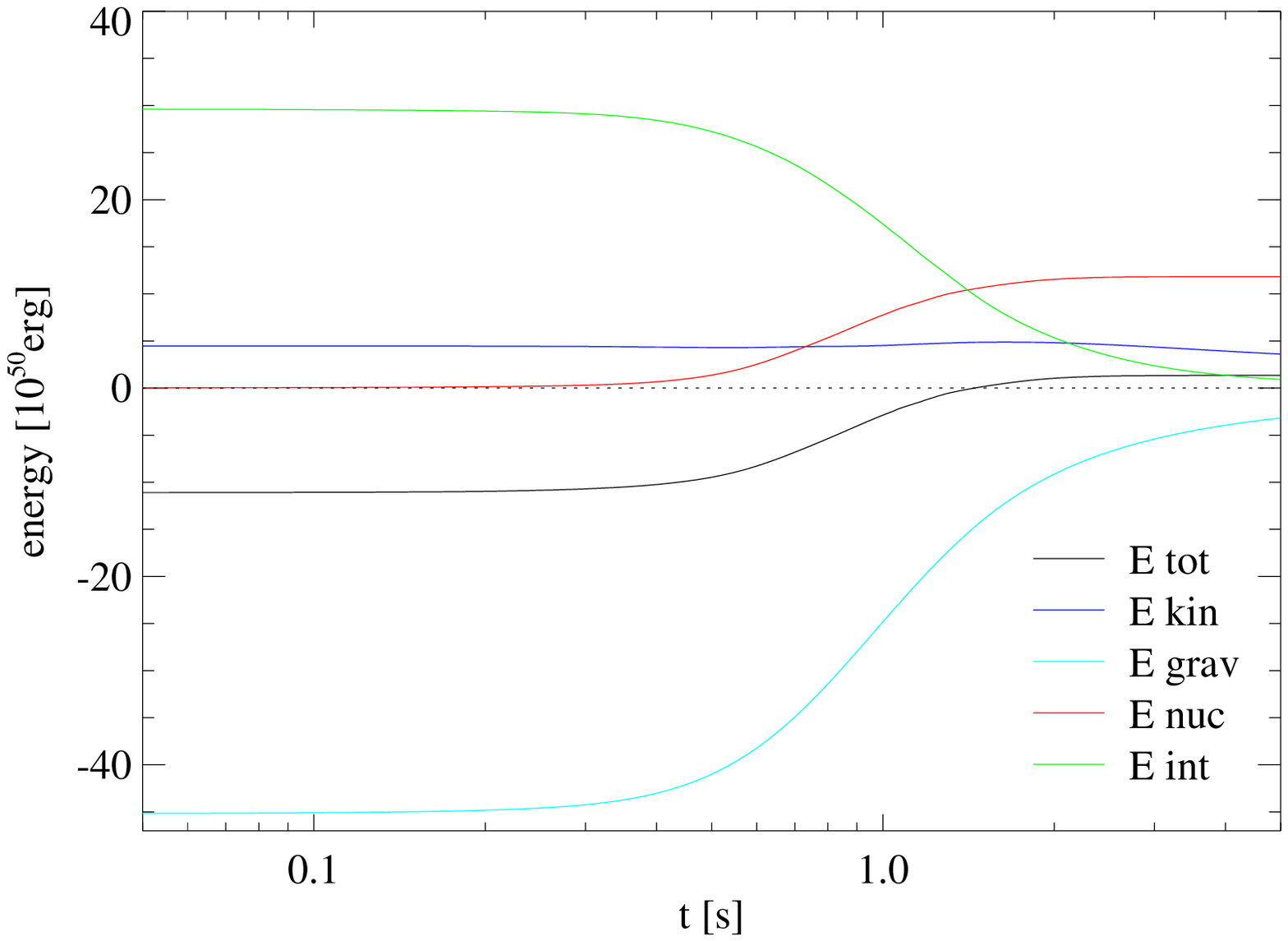}
      \label{fig:20AWD3_deflagration_eevol}}
  \end{minipage}\hfill
  \begin{minipage}{0.5\textwidth}
    \centering
  \subfigure[``AWD3 detonation'']
    {\includegraphics[width=\linewidth]{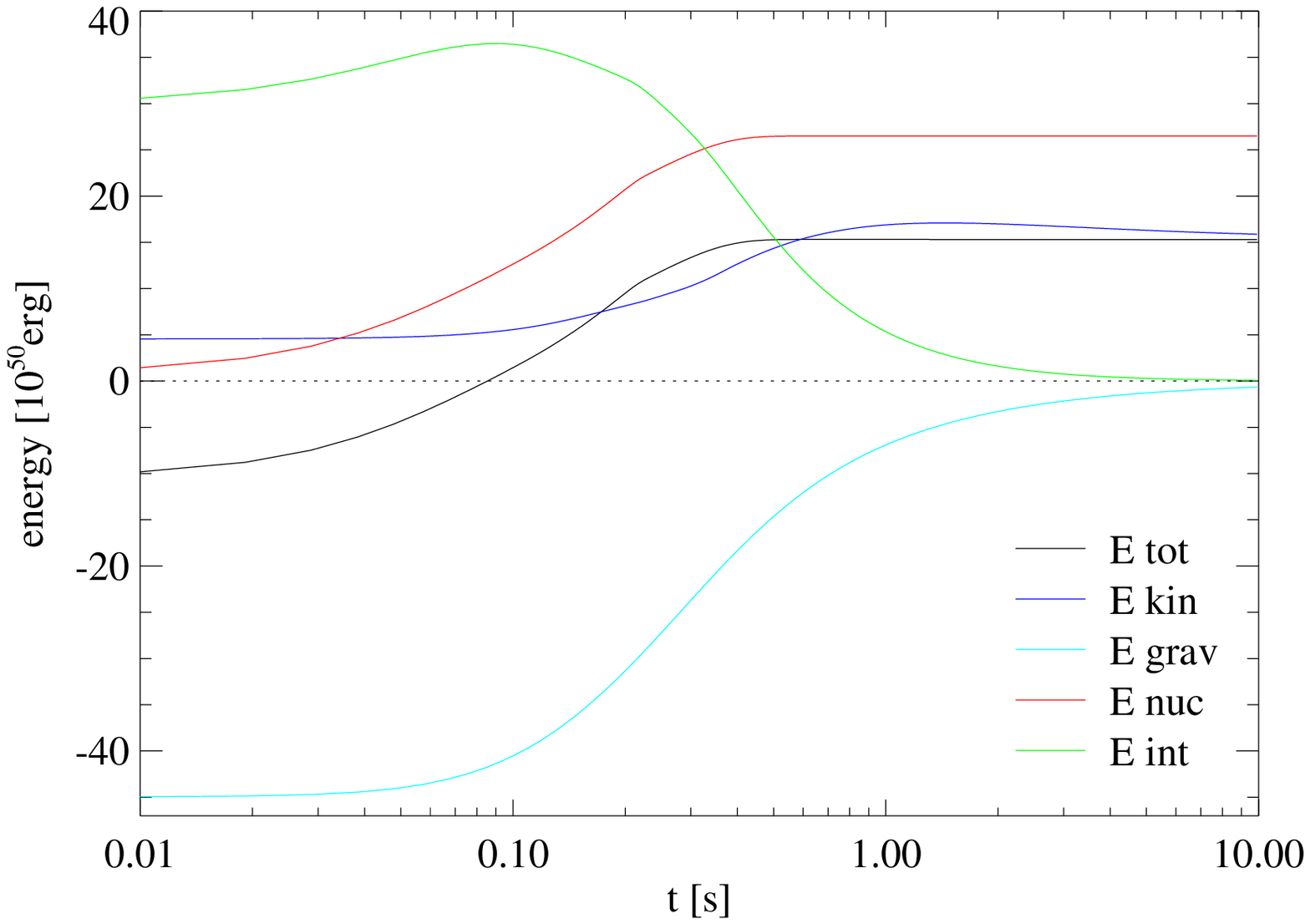}
      \label{fig:22AWD3_detonation_eevol}}
  \end{minipage}
  \caption{Evolution of the energy fractions for a deflagration
  scenario (a) and the prompt detonation, HBT limiting case (b).} 
  \label{fig:snpd_eevol_both}
\end{figure}

Compared to the subsonic deflagrations, the entire explosion proceeds
faster for detonations because of the supersonic propagation speed. 
Figure \ref{fig:snpd_eevol_both} shows the 
evolution of the energy contributions of the turbulent deflagration of
a similar rotator and the ``AWD3 detonation'' explosion model. The high
temperatures caused by the detonation lead to a noticeable peak in the
internal energy at $t \sim 0.1~\mathrm{s}$. The nuclear energy
released by the detonation exceeds $E_{nuc}$ for the deflagration scenario by
the 
factor of~$\sim 2.5$. This energy release is sufficient for an early unbinding of the star (which
is only marginally achieved in the deflagration case for the specific
deflagration scenario) and the ejection of material
with a kinetic energy that is increased by a factor of~$\sim
3$. 

As previously mentioned, the carbon-to-oxygen ratio was fixed to
X($^{12}\mathrm{C}) = 0.5$ for the detonation study. However, several
publications indicate that 
in the WD's interior it is less than 0.5
\citep{1975ApJ...196..791C,1999ApJ...513..861U, 
2002ApJ...568..779H,2006MNRAS.368..187L}, which would produce
a less energetic explosion.

%%%%%%%%%%%%%%%%%%%%%%%%%%%%%%%%%%%%%%%%%%%%%%%%%%%%%%%%%%%%%%%%%%%%%%%%%
%%%%%%%%%%%%%%%%%%%%%%%%%%%%%%%%%%%%%%%%%%%%%%%%%%%%%%%%%%%%%%%%%%%%%%%%%
%%%%%%%%%%%%%%%%%%%%%%%%%%%%%%%%%%%%%%%%%%%%%%%%%%%%%%%%%%%%%%%%%%%%%%%%%

\begin{figure*}[th]
\centering
  {\includegraphics[width=15cm]{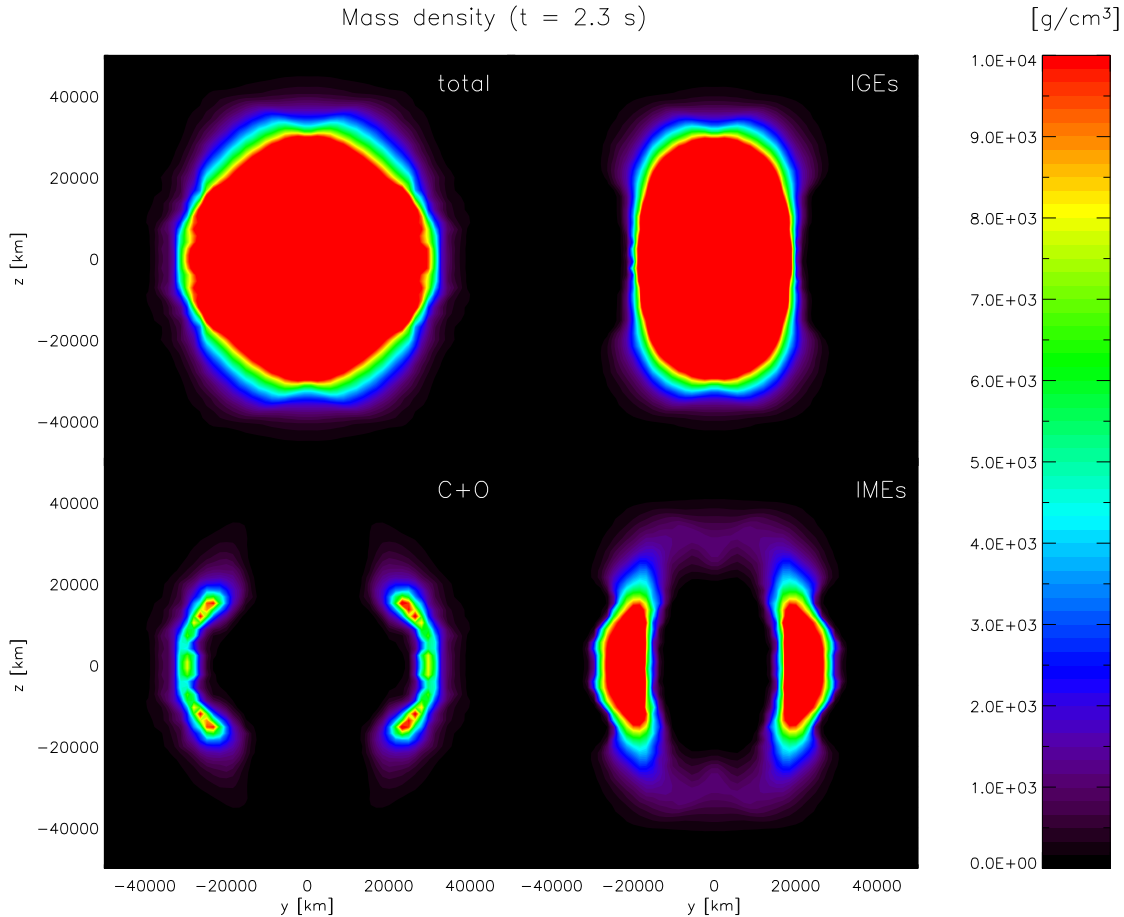}
    \label{fig:22AWD3_fracdens}}
  \caption{Total (upper left) as well as
    fractional mass densities (upper right: IGEs; lower left: unburnt fuel;
    lower right: IMEs) of the AWD3 rotator after $t = 2.3~\mathrm{s}$ (HBT
    limiting case). }
 \end{figure*}

\section{Expected spectral features}

The most striking property with respect to spectral features is the
extreme distribution of species in the homologous expansion
phase. Figure \ref{fig:22AWD3_fracdens} shows cross-sections along the rotation
axis for the total and fractional mass densities at $t =
2.3~\mathrm{s}$. The spatial extent of the ejecta at that time
corresponds approximately to the expansion reached for pure
deflagrations after $t = 5~\mathrm{s}$ (cf. Paper~I). Whereas IMEs are
present only in a torus within the equatorial plane, IGEs emerge close to the
poles and are therefore already visible at an early period. No fuel is
left at the centre of the star. The spatial distribution of the nuclear species should
produce a strong dependence on the line of sight. 

\begin{figure}[th]
  \begin{minipage}{0.5\textwidth}
    \centering
  \subfigure[``AWD3 deflagration $C_e = 5 \times 10^{\,3}\,$'']
    {\includegraphics[width=\linewidth]{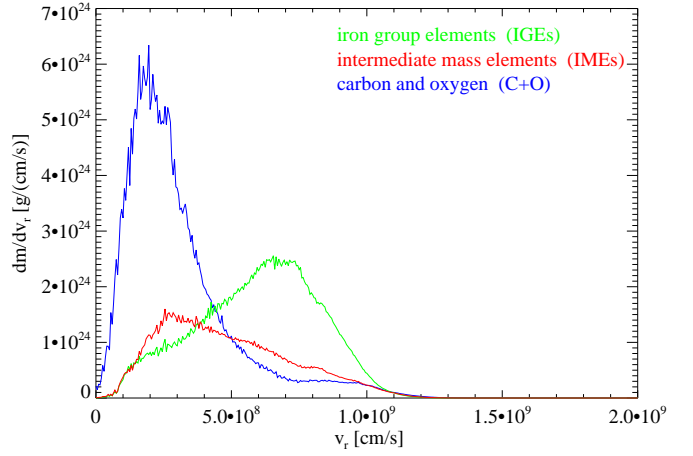}
      \label{fig:20AWD3_deflagration_massdist}}
  \end{minipage}\hfill
  \begin{minipage}{0.5\textwidth}
    \centering
   \subfigure[``AWD3 detonation'']
   {\includegraphics[width=\linewidth]{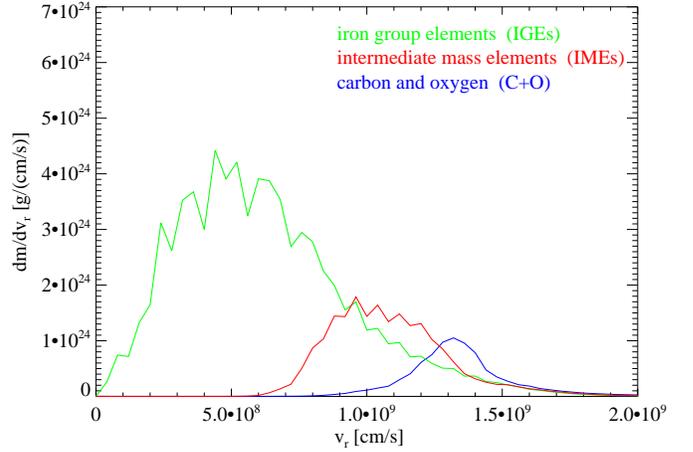}
      \label{fig:22AWD3_detonation_massdist}}
  \end{minipage}
  \caption{Probability density functions in radial velocity space for
  the deflagration scenario
  (a) and the prompt
  detonation (HBT limiting case) (b) at $t = 5~\mathrm{s}$. The 
  velocity increments are different for the deflagration and detonation
  simulation ($\mathrm{d}v = 5 \times 10^{\,6}~\mathrm{cm/s}$ and $\mathrm{d}v = 4
  \times 10^{\,7}~\mathrm{cm/s}$, respectively) as a result of the different
  resolution for both hydrodynamics simulations.}
  \label{fig:snpd_massdist}
\end{figure}

Figure \ref{fig:snpd_massdist} shows the resulting radial velocities at $t = 5~\mathrm{s}$ for the
different species in comparison to the deflagration model. No fuel
appears at low velocities, which is a necessary 
feature of an SN~Ia model. 
IGEs are encountered over a broad range of
velocities but are most prominent at lower radial velocities ($v_r \sim 5
\times 10^{\,3}~\mathrm{km/s}$). IMEs are found at higher
velocities, and are most prominent at $v_r \sim 10 \times
10^{\,3}~\mathrm{km/s}$. In addition, some of the ejecta exceed
velocities of up to $v_r \sim 20 \times
10^{\,3}~\mathrm{km/s}$. Therefore, high velocity features (HVFs; cf. 
\citet{2005ApJ...623L..37M}) could arise
from prompt 
detonations of rapidly rotating WDs. The composition structure and
velocity satisfy constraints that derived
for SN 1991T using a synthetic spectrum
\citep{1999MNRAS.304...67F}. Additional benefits of the prompt
detonation of the AWD3 rotator are given by work on the
three-dimensional spectral synthesis for early spectra
\citep{2006ApJ...645..470T}. It was found that a thick torus ---
similar to the one embodied by the IMEs --- can naturally explain the
diversity in observed strength of the HVFs by the covering of parts of the
star. 

%%%%%%%%%%%%%%%%%%%%%%%%%%%%%%%%%%%%%%%%%%%%%%%%%%%%%%%%%%%%%%%%%%%%%%%%%
%%%%%%%%%%%%%%%%%%%%%%%%%%%%%%%%%%%%%%%%%%%%%%%%%%%%%%%%%%%%%%%%%%%%%%%%%
%%%%%%%%%%%%%%%%%%%%%%%%%%%%%%%%%%%%%%%%%%%%%%%%%%%%%%%%%%%%%%%%%%%%%%%%%

\section{Post-processing of the prompt detonation}
\label{sc:post}

By considering a detailed nuclear reaction network,  
we gained insight into the composition
resulting from the ``AWD3 detonation'' scenario
by means of post-processing
(cf. \citet{2004A&A...425.1029T} for details of the method). 
In particular, we compared the composition obtained from
the post-processing to the composition 
resulting from the density-dependent burning of the hydrodynamical
simulation.
Even though differences in the derived compositions are discernable
(see
Table \ref{tab:yields_mod22}, in which all species
  exceeding the amount of $5 \times 10^{\,-3}~M_{\sun}$ are listed), {\it a significant amount of IMEs resulting
from the prompt detonation of an accreting WD rotator} is produced.

\begin{table}%[tp]
\begin{center}

\begin{tabular}{|l|c||l|c|}
\hline
IMEs                        & [M$_{\sun}$] & IGEs  &  [M$_{\sun}$] \\
\hline\hline 
$^{28}$Si           & 0.15 & $^{56}$Ni            & 1.48   \\
$^{32}$S            & 0.08 & $^{58}$Ni            & 0.13   \\
$^{36}$Ar           & 0.02 & $^{57}$Ni            & 0.05   \\
$^{40}$Ca           & 0.02 & $^{54}$Fe            & 0.05   \\
$^{24}$Mg           & 0.01 & $^{60}$Zn            & 0.01   \\
                    &      & $^{52}$Fe            & 0.01    \\
                    &      & $^{55}$Co            & 0.01    \\
                    &      & $^{62}$Zn            & 0.01    \\
\hline
\end{tabular}
\end{center}
\caption{Composition of various species for explosion scenario of
  ``AWD3
  detonation'' from the post-processing calculation.}
\label{tab:mod22_detail}
\end{table}

The detailed composition calculated by means of post-processing is listed in 
Table~\ref{tab:mod22_detail}. 
The total amount $0.29~M_{\sun}$
generated within the
AWD3 rotator
compares fairly well to the SN 1991T-like events, which exhibit more than a
solar mass of $^{56}\mathrm{Ni}$ and a small amount of $0.2$ to
$0.3~M_{\sun}$ of IMEs (P. Mazzali, private communication). Table
\ref{tab:mod22_detail} 
lists all the isotopes of IGEs and IMEs that exceed five-tenths of a
percent of a solar mass. The amount of $1.48~M_{\sun}$ of
$^{56}\mathrm{Ni}$ is extraordinarily high and should lead to very bright
SNe~Ia. As a result of
high temperatures, fewer 
neutron-rich isotopes among IGEs are produced. For instance, the 
ratio of $^{56}\mathrm{Ni} \,/\,^{58}\mathrm{Ni} = 11$.  Whereas the
radioactive decay 
of $^{56}\mathrm{Ni}$ supplies the energy for the light emitted by the
SN~Ia ejecta, the presence of other iron-group nuclei enhances the
opacity of the ejecta and causes a broadening of the light curve
\citep{2002A&A...391.1167R}. 

%%%%%%%%%%%%%%%%%%%%%%%%%%%%%%%%%%%%%%%%%%%%%%%%%%%%%%%%%%%%%%%%%%%%%%%%%
%%%%%%%%%%%%%%%%%%%%%%%%%%%%%%%%%%%%%%%%%%%%%%%%%%%%%%%%%%%%%%%%%%%%%%%%%
%%%%%%%%%%%%%%%%%%%%%%%%%%%%%%%%%%%%%%%%%%%%%%%%%%%%%%%%%%%%%%%%%%%%%%%%%

\section{Detonating rotators: only an exotic scenario?}

We have demonstrated that the prompt detonation of a
rapidly rotating WD potentially meets the criteria of 
superluminous SNe~Ia regarding spectral and bolometric features. 
In particular, prompt detonations are capable of  producing a sufficient
amount of IMEs, if the progenitor star is a super-Chandrasekhar-mass 
WD with differential rotation as proposed by \citet{2005A&A...435..967Y}.
Therefore, this explosion scenario may provide an explanation of the rather small number
of events such as SN~1991T or SN 2003fg. 

In our investigation, we assumed that prompt detonations can occur in 
rapidly rotating WDs. For WDs that do not rotate appreciably, on the
other hand, pure deflagration or delayed detonation models, where the
DDT occurs at a relatively 
late stage of the explosion, are most successful. The prompt
detonation may be seen as a limiting case of a
deflagration-to-detonation transition, where the transition time
approaches zero as the rotation frequency of the WD becomes nearly
critical. But this appears to be implausible, if the
mechanism causing DDTs is related to the intermittency of turbulence
\citep{2008ApJ...681..470P,SchmCir09}. 
Alternatively, prompt detonations might occur beyond a certain
rotation frequency, whereas 
detonations are delayed or entirely absent for WDs rotating at lower
frequencies. However, given the rudimentary theoretical understanding
of the onset of explosive thermonuclear burning in WDs (both 
non-rotating and rotating), no mechanism that selects the initial burning mode depending
on the rotation frequency is known at present. 

\begin{acknowledgements} 
We thank Fritz R\"{o}pke for post-processing the 
``AWD3 detonation'' explosion model, Paolo Mazzali for discussing the
results, and the referee for helpful remarks about the NSE transition
densities. 

\end{acknowledgements}

\bibliographystyle{aa}
\bibliography{12033_colour}

\begin{thebibliography}{37}
\expandafter\ifx\csname natexlab\endcsname\relax\def\natexlab#1{#1}\fi

\bibitem[{{Arnett}(1969)}]{Arn69}
{Arnett}, D.~W. 1969, \apss, 5, 180

\bibitem[{{Couch} \& {Arnett}(1975)}]{1975ApJ...196..791C}
{Couch}, R.~G. \& {Arnett}, W.~D. 1975, \apj, 196, 791

\bibitem[{{Fisher} {et~al.}(1999){Fisher}, {Branch}, {Hatano}, \&
  {Baron}}]{1999MNRAS.304...67F}
{Fisher}, A., {Branch}, D., {Hatano}, K., \& {Baron}, E. 1999, \mnras, 304, 67

\bibitem[{{Gamezo} {et~al.}(2005){Gamezo}, {Khokhlov}, \& {Oran}}]{GamKhok05}
{Gamezo}, V.~N., {Khokhlov}, A.~M., \& {Oran}, E.~S. 2005, \apj, 623, 337

\bibitem[{{Gamezo} {et~al.}(1999){Gamezo}, {Wheeler}, {Khokhlov}, \&
  {Oran}}]{Gamezoetal1999}
{Gamezo}, V.~N., {Wheeler}, J.~C., {Khokhlov}, A.~M., \& {Oran}, E.~S. 1999,
  \apj, 512, 827

\bibitem[{{Golombek} \& {Niemeyer}(2005)}]{2005A&A...438..611G}
{Golombek}, I. \& {Niemeyer}, J.~C. 2005, \aap, 438, 611

\bibitem[{{H\"oflich} \& {Stein}(2002)}]{2002ApJ...568..779H}
{H\"oflich}, P. \& {Stein}, J. 2002, \apj, 568, 779

\bibitem[{{Howell} {et~al.}(2006){Howell}, {Sullivan}, {Nugent}, {Ellis},
  {Conley}, {Le Borgne}, {Carlberg}, {Guy}, {Balam}, {Basa}, {Fouchez}, {Hook},
  {Hsiao}, {Neill}, {Pain}, {Perrett}, \& {Pritchet}}]{Howell_etal}
{Howell}, D.~A., {Sullivan}, M., {Nugent}, P.~E., {et~al.} 2006, \nat, 443, 308

\bibitem[{{Imshennik} \& {Khokhlov}(1984)}]{ImshennikKhokhlov1984}
{Imshennik}, V.~S. \& {Khokhlov}, A.~M. 1984, Pis ma Astronomicheskii Zhurnal,
  10, 631

\bibitem[{{Jeffery} {et~al.}(2006){Jeffery}, {Branch}, \&
  {Baron}}]{Jeffery_etal}
{Jeffery}, D.~J., {Branch}, D., \& {Baron}, E. 2006, eprint
  ArXiv:astro-ph/0609804

\bibitem[{{Kasliwal} {et~al.}(2008){Kasliwal}, {Ofek}, {Gal-Yam}, {Rau}, \&
  {Brown}}]{Kasliwal_etal}
{Kasliwal}, M.~M., {Ofek}, E.~O., {Gal-Yam}, A., {Rau}, A., \& {Brown}, P.~J.
  e.~a. 2008, \apjl, 683, L29

\bibitem[{{Khokhlov}(1991)}]{1991A&A...245..114K}
{Khokhlov}, A.~M. 1991, \aap, 245, 114

\bibitem[{{Lesaffre} {et~al.}(2006){Lesaffre}, {Han}, {Tout}, {Podsiadlowski},
  \& {Martin}}]{2006MNRAS.368..187L}
{Lesaffre}, P., {Han}, Z., {Tout}, C.~A., {Podsiadlowski}, P., \& {Martin},
  R.~G. 2006, \mnras, 368, 187

\bibitem[{{Mazzali} {et~al.}(2005){Mazzali}, {Benetti}, {Altavilla}, {Blanc},
  {Cappellaro}, {Elias-Rosa}, {Garavini}, {Goobar}, {Harutyunyan}, {Kotak},
  {Leibundgut}, {Lundqvist}, {Mattila}, {Mendez}, {Nobili}, {Pain},
  {Pastorello}, {Patat}, {Pignata}, {Podsiadlowski}, {Ruiz-Lapuente}, {Salvo},
  {Schmidt}, {Sollerman}, {Stanishev}, {Stehle}, {Tout}, {Turatto}, \&
  {Hillebrandt}}]{2005ApJ...623L..37M}
{Mazzali}, P.~A., {Benetti}, S., {Altavilla}, G., {et~al.} 2005, \apjl, 623,
  L37

\bibitem[{{Niemeyer}(1999)}]{1999ApJ...523L..57N}
{Niemeyer}, J.~C. 1999, \apjl, 523, L57

\bibitem[{{Osher} \& {Sethian}(1988)}]{OshSet88}
{Osher}, S. \& {Sethian}, J.~A. 1988, Journal of Computational Physics, 79, 12

\bibitem[{{Pan} {et~al.}(2008){Pan}, {Wheeler}, \&
  {Scalo}}]{2008ApJ...681..470P}
{Pan}, L., {Wheeler}, J.~C., \& {Scalo}, J. 2008, \apj, 681, 470

\bibitem[{{Pfannes} {et~al.}(2009){Pfannes}, {Niemeyer}, {Schmidt}, \&
  Klingenberg}]{PaperI}
{Pfannes}, J.~M.~M., {Niemeyer}, J.~C., {Schmidt}, W., \& Klingenberg, C. 2009,
  \aap, submitted

\bibitem[{{Plewa} {et~al.}(2004){Plewa}, {Calder}, \& {Lamb}}]{GCDmodel}
{Plewa}, T., {Calder}, A.~C., \& {Lamb}, D.~Q. 2004, \apjl, 612, L37

\bibitem[{Reinecke(2001)}]{Rei01}
Reinecke, M. 2001, PhD thesis, MPA Garching

\bibitem[{{Reinecke} {et~al.}(2002){Reinecke}, {Hillebrandt}, \&
  {Niemeyer}}]{2002A&A...391.1167R}
{Reinecke}, M., {Hillebrandt}, W., \& {Niemeyer}, J.~C. 2002, \aap, 391, 1167

\bibitem[{{Reinecke} {et~al.}(1999){Reinecke}, {Hillebrandt}, {Niemeyer},
  {Klein}, \& {Gr\"{o}bl}}]{RHNKG99b}
{Reinecke}, M., {Hillebrandt}, W., {Niemeyer}, J.~C., {Klein}, R., \&
  {Gr\"{o}bl}, A. 1999, \aap, 347, 724

\bibitem[{{R{\"o}pke} {et~al.}(2007){R{\"o}pke}, {Hillebrandt}, {Schmidt},
  {Niemeyer}, {Blinnikov}, \& {Mazzali}}]{snob}
{R{\"o}pke}, F.~K., {Hillebrandt}, W., {Schmidt}, W., {et~al.} 2007, \apj, 668,
  1132

\bibitem[{{R{\"o}pke} \& {Niemeyer}(2007)}]{RoepkeNiemeyer07}
{R{\"o}pke}, F.~K. \& {Niemeyer}, J.~C. 2007, \aap, 464, 683

\bibitem[{{Schmidt} {et~al.}(2009){Schmidt}, {Ciaraldi-Schoolmann}, {Niemeyer},
  {Roepke}, \& {Hillebrandt}}]{SchmCir09}
{Schmidt}, W., {Ciaraldi-Schoolmann}, F., {Niemeyer}, J.~C., {Roepke}, F.~K.,
  \& {Hillebrandt}, W. 2009, \apj, submitted

\bibitem[{{Schmidt} \& {Niemeyer}(2006)}]{2006A&A...446..627S}
{Schmidt}, W. \& {Niemeyer}, J.~C. 2006, \aap, 446, 627

\bibitem[{{Sharpe}(1999)}]{Sha99}
{Sharpe}, G.~J. 1999, \mnras, 310, 1039

\bibitem[{{Stanishev} {et~al.}(2007){Stanishev}, {Taubenberger}, {Blanc},
  {Anupama}, \& {Benetti}}]{Stanishev_etal}
{Stanishev}, V., {Taubenberger}, S., {Blanc}, G., {Anupama}, G.~C., \&
  {Benetti}, S. e.~a. 2007, in American Institute of Physics Conference Series,
  Vol. 924, The Multicolored Landscape of Compact Objects and Their Explosive
  Origins, ed. T.~{di Salvo}, G.~L. {Israel}, L.~{Piersant}, L.~{Burderi},
  G.~{Matt}, A.~{Tornambe}, \& M.~T. {Menna}, 336--341

\bibitem[{{Steinmetz} {et~al.}(1992){Steinmetz}, {M\"uller}, \&
  {Hillebrandt}}]{1992A&A...254..177S}
{Steinmetz}, M., {M\"uller}, E., \& {Hillebrandt}, W. 1992, \aap, 254, 177

\bibitem[{{Tanaka} {et~al.}(2006){Tanaka}, {Mazzali}, {Maeda}, \&
  {Nomoto}}]{2006ApJ...645..470T}
{Tanaka}, M., {Mazzali}, P.~A., {Maeda}, K., \& {Nomoto}, K. 2006, \apj, 645,
  470

\bibitem[{{Timmes} {et~al.}(2000){Timmes}, {Zingale}, {Olson}, {Fryxell},
  {Ricker}, {Calder}, {Dursi}, {Tufo}, {MacNeice}, {Truran}, \&
  {Rosner}}]{Timmesetal2000}
{Timmes}, F.~X., {Zingale}, M., {Olson}, K., {et~al.} 2000, \apj, 543, 938

\bibitem[{{Travaglio} {et~al.}(2004){Travaglio}, {Hillebrandt}, {Reinecke}, \&
  {Thielemann}}]{2004A&A...425.1029T}
{Travaglio}, C., {Hillebrandt}, W., {Reinecke}, M., \& {Thielemann}, F.-K.
  2004, \aap, 425, 1029

\bibitem[{{Umeda} {et~al.}(1999){Umeda}, {Nomoto}, {Yamaoka}, \&
  {Wanajo}}]{1999ApJ...513..861U}
{Umeda}, H., {Nomoto}, K., {Yamaoka}, H., \& {Wanajo}, S. 1999, \apj, 513, 861

\bibitem[{{Woosley} {et~al.}(2008){Woosley}, {Kerstein}, {Sankaran}, \&
  {Roepke}}]{WoosKer09}
{Woosley}, S.~E., {Kerstein}, A.~R., {Sankaran}, V., \& {Roepke}, F.~K. 2008,
  \apj, submitted, eprint arXiv:0811.3610

\bibitem[{{Woosley} \& {Weaver}(1994)}]{WW94}
{Woosley}, S.~E. \& {Weaver}, T.~A. 1994, in Les Houches Session LIV, ed. S.~A.
  {Bludman}, R.~{Mochkovitch}, \& J.~{Zinn-Justin}, North-Holland

\bibitem[{{Yoon} \& {Langer}(2004)}]{2004A&A...419..623Y}
{Yoon}, S.-C. \& {Langer}, N. 2004, \aap, 419, 623

\bibitem[{{Yoon} \& {Langer}(2005)}]{2005A&A...435..967Y}
{Yoon}, S.-C. \& {Langer}, N. 2005, \aap, 435, 967

\end{thebibliography}

\end{document}